\documentclass[11pt,a4paper]{article}
\usepackage{graphicx}
\newcommand{\bery}{$^{7}\mbox{Be}$}
\author{Zhiyi LIU, Chengbo LI, Siguang WANG, Jing ZHOU, \\Qiuying MENG, Shaojun LU, Shuhua ZHOU}
\title{Measurement of Change of $^{7}\mbox{Be}$ decay rate \\in Be and Au}

\begin{document}

\maketitle
\begin{abstract} \textit{We have measured the possible change of the decay rate of $^{\scriptsize{\textit{7}}}Be$ implanted
into hosts of natural beryllium and natural gold. No difference between the
$^{\scriptsize{\textit{7}}}Be$ decay rates in the two hosts is observed within the experimental
precision of 0.12\%. This result implies that change of the decay rate of
$^{\scriptsize{\textit{7}}}Be$ implanted in different materials cannot be simply expected from the
electron affinity difference consideration lonely and the lattice structure of the host materials
should be taken into account.}
\end{abstract}

Electron-capture decay rates depend sensitively on the density of atomic electrons within the
nucleus. Thus, physically and chemically environmental factors such as pressure, chemical form,
magnetic fields, etc. that can alter electron densities, may affect electron-capture decay rates.
Since the change of nuclear decay in different environments has fundamental significance, as well
as application in nuclear physics, geology and condensed matter physics, such a study is of
current interests.$^{\scriptsize{\cite{72Eme,77Dos,96Bah,99Ker,70Joh}}}$ \bery{} is the lightest
radioactive nucleus that decays by electron capture with a half-life of $\sim$ 53 days, thereby,
it is a good candidate for studying perturbation of nuclear decay rates. Furthermore, the study of
decay rate changes for \bery{} has particular significance to the solar neutrino problem, where
there is a large discrepancy between theoretical predictions and experimental determinations of
the solar neutrino flux.$^{\scriptsize{\cite{96Bah,97Wol,97Cha,02Jun}}}$

The \bery{} decay rate sensitively depends on chemical environment at the nucleus. Changes of
decay rates of \bery{} nuclei in different \bery{} compounds have been measured at normal pressure
and a maximum change is about 1.5\%.$^{\scriptsize{\cite{99Huh}}}$ Moreover, this decay rate is
also sensitive to physical environments, such as high pressure and host materials in which the
\bery{} nucleus is located. Liu et al. have measured a large increase, up to 1\%, of the decay
rate of \bery{} under a high pressure of 400 kilobars.$^{\scriptsize{\cite{00Liu}}}$ Recently
Norman et al. measured the decay rate of \bery{} implanted into hosts of lithium fluoride, gold,
graphite, boron nitride, tantalum, lithium, and so on. It has been found that the \bery{} decay
rate varies by as much as 0.72\% from one host to
another.$^{\scriptsize{\cite{96Jae,99Ray,01Nor,02Sou}}}$ Qualitatively, it has been reported that
if a \bery{} atom is implanted in a medium having high electron affinity (EA), as a result of its
interaction with nearby atoms of such a medium, the \bery{} atom would lose a significant fraction
of its 2\emph{s} electrons.$^{\scriptsize{\cite{99Ray}}}$ The decay rate of \bery{} in a high-EA
medium is thereby smaller than that in a low-EA medium. Thus, we would expect that \bery{}
implanted in natural gold (EA$_{\scriptsize{\mbox{Au}}}$=2.308eV $^{\scriptsize{\cite{02Lid}}}$)
should decay slower up to 0.7\% than that in natural beryllium
(EA$_{\scriptsize{\mbox{Be}}}$=гн0.19eV $^{\scriptsize{\cite{02Lid}}}$) according to the result of
Ray et al., in which difference up to 0.72\% between the decay rates of \bery{} implanted in Au
and in $\mbox{Al}_{2}\mbox{O}_{3}$ was observed.$^{\scriptsize{\cite{99Ray}}}$

In this Letter, we report our measurement of possible change of the decay rate of \bery{}
implanted into metal foils of natural beryllium and natural gold. The \bery{} nuclei are produced
by bombarding a 500 $\mu\mbox{g}/\mbox{cm}^{2}$-thick foil of lithium fluoride (LiF) with a 3.2
MeV proton beam, with average current 5 $\mu$A, from the 5SDH-2 tandem accelerator at CNNC
Radiation Metrology and Measurement Center, China Institute of Atomic Energy. Thus, \bery{} nuclei
produced by the reaction $^{7}$Li (p, n) \bery{} with recoil energy 1.0 MeV in the forward
direction are implanted into beryllium and gold foils placed immediately behind the LiF target.
The implantation for each foil lasts about 15 h. As a result of such implantations, \bery{} atoms
are expected to be randomly located in the interstitial lattice space of the host media beryllium
and gold.

By electron capture the nucleus \bery{} decays to the 3/2$^{-}$ ground state of $^{7}$Li directly
with a branching ratio of 89.5\% and to the first excited state with a ratio of 10.5\%, which
decays subsequently to its ground state by emitting a 478-keV gamma
ray.$^{\scriptsize{\cite{96Fir}}}$ In this experiment, two high-purity coaxial germanium (HP Ge)
detectors are used to measure the 478-keV gamma-ray photons. The \bery{} implanted beryllium and
gold foils are separately mounted about 2 cm away from the endcups of two HP Ge detectors. The two
detectors, each is surrounded by a graded shield of organic glass-Al-Cu-Pb, are located four
meters away from each other to avoid cross detection of the gamma-ray photons from the two
sources. The gamma-ray spectra are accumulated with a KODAQ data acquisition system.

\begin{figure}[h]
 \begin{center}
  \begin{minipage}[b]{10cm}
  \includegraphics[width=80mm]{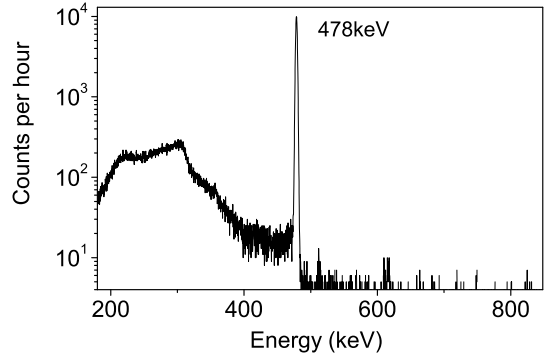}
  \caption{\label{spec}A typical gamma-ray spectrum obtained in one-hour counting interval from \bery{} in the gold sample}
  \end{minipage}
 \end{center}
\end{figure}

In order to decrease as much effect originated from the decays of short-lived isotopes produced in
bombarding as possible, we wait for about 40 days to start to acquire experimental data. The two
gamma-ray spectra are acquired for successive one-hour intervals and recorded on a computer hard
disk. Then this is followed by next interval automatically. After such a measurement has lasted
about 70 days, we exchange the positions of the two sources, while keeping other measurement
conditions unchanged, and we measure for another 50 days. We extract from each time bin the net
peak area of the 478 keV  \bery{} gamma-ray and its time information after dead-time correction
and time calibrations with the time provided by Hongkong observatory. Figure \ref{spec} shows a
typical gamma-ray spectrum in a one-hour counting period from the gold sample. The ratio method,
as to be described below, is employed to draw the possible difference in the decay rate of \bery{}
nuclei implanted in beryllium and gold. Let $A_{Be}$ and $A_{Au}$ denote the numbers of 478 keV
gamma-ray photons measured by the corresponding detectors within an interval of $t$ to ($t$+1)
days. Similarly, let $A^0_{Be}$ and $A^0_{Au}$ denote the numbers of 478 keV gamma-ray photons
measured by the corresponding detectors within an interval of 0 to $t$=1 days. Let $\lambda_{Be}$
and $\lambda_{Au}$ denote the decay rate of the \bery{} nuclei in the beryllium and gold samples
respectively and let $\lambda_{Be}=\lambda_{Au}+\Delta\lambda$. Then
\begin{equation} \mbox{ln}R(t)=\mbox{ln}R_{0}-\Delta\lambda t
\end{equation}
where $R(t)=\frac{A_{Be}}{A_{Au}}$, $R_0=\frac{A^0_{Be}}{A^0_{Au}}$, and $\Delta\lambda$ is the
difference in the decay rate.
\begin{figure}[h]
 \begin{center}
  \begin{minipage}[t]{10cm}
  \includegraphics[width=80mm]{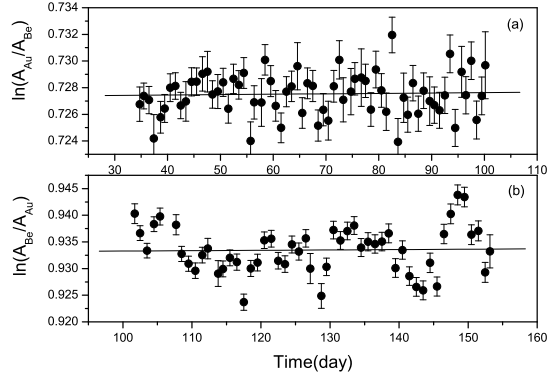}
  \caption{\label{ln-t}Characteristics of $\mbox{ln}R$ versus time (a) before and (b) after exchanging sources positions. The solid lines are the linear fit of the experimental data.}
  \end{minipage}
 \end{center}
\end{figure}

Figures \ref{ln-t}(a) and \ref{ln-t}(b) show the experimental data of $\mbox{ln}R(t)$ versus $t$
and corresponding linear fit lines before and after exchanging the source positions, respectively.
The $\Delta\lambda$ value is determined by the slope in each case. The value of $\Delta\lambda$
derived from Figs. 2(a) and 2(b) are $0.3\times10^{-5}\pm0.8\times10^{-5}$ and
$-0.7\times10^{-5}\pm1.5\times10^{-5}$, respectively. By fitting the exponential decrease of
$A_{Au}$ versus time $t$ for the two cases, we obtain the weighted value $0.012985\pm0.000004$ for
the decay rate of \bery{} implanted in gold and $0.012986\pm0.000004$ in beryllium. Then using
$\lambda$ equal to 0.012986, we obtain the values of $(0.02\pm0.06)$\% before the source exchange
and ($-0.05\pm0.12$)\% after the source exchange. This result indicates that within our
experimental precision, the large change in the decay rate of \bery{} in Be and Au is not observed
and an upper limit 0.12\% of this change can be set.

In conclusion, we have performed the first-ever measurement of the possible change in the decay
rate of \bery{} implanted in natural beryllium and natural gold. According to the large difference
of electron affinity between beryllium and gold, a large change of 0.7\% in the decay rate of
\bery{} in the two cases is expected. However, within our measurement precision ($\sim\pm0.06$ and
0.12\% for the measurement before and after the source exchange respectively) no difference
between the decay rates of \bery{} implanted in beryllium and in gold is observed. This result
implies that except electron affinity the lattice structure of the host medium in which the
\bery{} atom sits has to be taken into account.

Our experimental result also indicates that the difference between the average electron numbers in
$L$-shell (2$s$ state) of \bery atoms in the two samples is smaller than our measurement precision
of $\sim0.1$\%, since the $K$-shell capture rate should essentially remain unchanged in different
environments. According to the calculation of Ray et al., which has taken the lattice structure
into account, the average number of 2$s$ electrons of \bery{} atoms implanted in natural beryllium
sample should be equal to 0.443 while that in the gold sample should be
0.416.$^{\scriptsize{\cite{99Ray,02Ray01,02Ray02}}}$ Thus, from Hartree's calculation that the
ratio of the square of the beryllium 2$s$ electronic state wave function (2$s$ electrons) to that
of 1$s$ state wave function at the nucleus (r=0) is 3.31\%,$^{\scriptsize{\cite{35Har}}}$ the
decay rate of \bery{} in natural beryllium should be faster than that in gold by 0.045\%. However,
this value is beyond the reach of our measurement precision.
\subsection*{Acknowledgement}We are grateful to Amlan Ray at Variable Energy Cyclotron Centre of India for useful discussion.
We also acknowledge Da-Qing Yuan and Chao-Fan Rong for using their gamma-ray spectroscopy
laboratory.


\end{document}